# Enhancing Evacuation Planning through Multi-Agent Simulation and Artificial Intelligence: Understanding Human Behavior in Hazardous Environments

**Afnan Alazbah[1], Khalid Fakeeh[2], Osama Rabie[3]**
[1]aazbah0001@stu.kau.edu.sa, [2]kfakeeh@kau.edu.sa, [3]obrabie@kau.edu.sa
[1, 2, 3]Information Systems Department, King Abdul-Aziz University, Jeddah, Saudi Arabia

*Abstract*— **This paper focuses on the crucial task of addressing the evacuation of hazardous places, which holds great importance for coordinators, event hosts, and authorities. To facilitate the development of effective solutions, the paper employs Artificial Intelligence (AI) techniques, specifically Multi-Agent Systems (MAS), to construct a simulation model for evacuation. NetLogo is selected as the simulation tool of choice due to its ability to provide a comprehensive understanding of human behaviour in distressing situations within hazardous environments. The primary objective of this paper is to enhance our comprehension of how individuals react and respond during such distressing situations. By leveraging AI and MAS, the simulation model aims to capture the complex dynamics of evacuation scenarios, enabling policymakers and emergency planners to make informed decisions and implement more efficient and effective evacuation strategies. This paper endeavours to contribute to the advancement of evacuation planning and ultimately improve the safety and well-being of individuals in hazardous places.**

## I. INTRODUCTION

Since the first times that individuals were gathered in small spaces, it has been clear that distressing situations, like emergencies, call for effective evacuation. The safety of individuals in enclosed places is currently seriously threatened by several dangerous events, but happily, there are several ways to reduce the potential harm. The primary objective of this paper is to construct an agent-based model that simulates the evacuation of military and civilians in enclosed places while also taking into account the effect of their emotional condition on those nearby. It takes time and effort to develop the tools required to tackle an issue this complex. The construction of our simulation model, however, becomes more flexible and effective by utilising already-existing technologies. Another objective of this paper is to use software simulation tools to build a ubiquitous MAS made up of intelligent software agents. These agents specifically should be able to "*infect*" other agents close to the system boundaries with contagious emotions. The agents should also be able to simulate the whole spectrum of emotions that real individuals experience. A model like this might be used by experts in a variety of industries to research how individuals behave under stress and use their findings to improve the architecture of public venues like lecture halls and theatres. The ultimate objective is to create a thorough simulation model that faithfully reproduces the conditions and dynamics seen in hazardous places while taking into account the various backgrounds, personalities, moods, and emotional states of the individuals. Several objectives must be met to accomplish the paper's objective which are described as follows:

1. Learn more about how personality, mood, and emotions affect how individuals behave in life-threatening situations like panic and awareness.

2. Build up the knowledge of MAS to the point that readers can use them well.

3. The ability to use the NetLogo simulation platform at a proficient level to create a multi-agent evacuation simulation model with the knowledge learned in objective 1. Then, showcasing the required features in a simulation that represents the final software implementation.

The paper is as follows: The background of the agents that will be employed in this paper is covered in the following section. The related works that are closely connected to our methodology have been presented in Section III. Early experiments and a description of the model are included in Section IV of the paper. The later experiments, together with the tools employed and how they were implemented, are discussed in Section V. The NetLogo simulations are shown in Section VI. The simulation results have been given under various scenarios in Section VII, and in Section VIII, we draw some conclusions and discuss possible future research.

## II. BACKGROUND

In this section, evaluate and distinguish the cognitive science components that are pertinent to the paper. It seeks to give readers a thorough grasp and presentation of the knowledge and skills required to finish the work successfully. Although there are other kinds of agents, it's crucial to remember that this paper only focuses on a certain group of agents. Other types of agents are not mentioned or taken into account in this paper because they are outside its purview.

## A. Multi-Agent Systems

In relation to the idea of a MAS, there are numerous definitions available. The definition provided in [1] is more appropriate for the context of this paper, though. The study of group behaviour in live organisms through artificial life and Distributed Artificial Intelligence (DAI) are two places that contributed to the development of the MAS framework [2]. In other words, MAS is an area of AI that focuses on developing guidelines for creating complex models with several agents. Utilising complex techniques, these concepts allow autonomous agents to coordinate [3]. With regard to manufacturing MAS, there are a variety of observable facets of the universe that might be considered. The ubiquitous MAS is demonstrated, for instance, by the communication and energy propagation between the nodes of a Dyson Sphere that uses energy from a star within a solar system. In this case, a leading drone with sophisticated skills would approach the target star and create a virtual network using blueprints that were already in place. The formation started by the leading drone would then be completed by a swarm of follower drones communicating in coordination while conforming to anti-collision protocols. Another compelling example of MAS is demonstrated by the collective behaviour of a swarm of fish, functioning as a cohesive unit. Particularly, their attempts to promote species longevity are demonstrated by the reproduction cycle seen in the red handfish species, Thymichthys politics [4]. Although little is known about the red handfish, it has been noted that they are dependent on a particular, rare plant to deposit their eggs. The cooperative quest by red handfish for these elusive nesting places could be depicted by MAS.

## B. Emotional Agent

This agent, which is derived from the multidisciplinary field of cognitive science, focuses on integrating emotions into intelligent agents. Agents in MAS and real-world individuals are both directly impacted by several variables. These variables include emotional contagion, personality, assessment, and more [5], [6]. When emotions are incorporated into agents, the system becomes more difficult from a computer science standpoint [3]. Let's move on by giving succinct explanations of what qualifies an agent as an emotional agent:

- **Personality:** The core elements that drive personality, namely Openness, Consciousness, Extraversion, Neuroticism, and Agreeableness (OCEAN), can be used to determine a person's personality. The personality trait of the agent is initialised in the model with a fixed value that does not change.
- **Moods:** Moods are frequently wrongly equated with emotions. They are distinct from one another, though, in that moods are weaker, more diffuse, and last for a longer duration. Moods can have a big impact on individuals during an evacuation.

## III. RELATED WORKS

Generally, crowd simulation involves path planning, anti-collision, and crowd rendering in computer graphics, transportation science, and virtual reality. It is believed that it can be divided into micro-modelling and macro-modelling techniques. Micro-modelling techniques use agent-based models, including rule-based, force-based, velocity-based, and particle-based models. Agents act independently, with global movements guided by path planning and local interactions managed through collision avoidance. Examples include boid model [14], Social Force Model (SFM) [15], [16], and particle swarm optimization model [17]. Macro-modelling techniques treat the entire crowd as a continuous entity, inspired by fluid dynamics. It focuses on large-scale simulations without individual behaviour details. Treuille et al. [18] real-time simulation uses continuum dynamics, while Hughes [19] applies partial differential equations. Other than this, several researchers have explored realistic and adaptive crowd simulations such as O'Connor et al. [20] model incorporates perceived realism with social forces. Sung et al. [21] model achieves complex behaviours without increasing agent complexity. Taffou et al. [22] uses virtual reality to study group perception, and other studies modify virtual animations.

Researchers have utilized computer graphics, virtual reality, and game technology to develop intelligent emergency evacuation evaluation models such as Bellomo et al. [23] proposed a mesoscopic model for behavioural crowds in unbounded domains, focusing on understanding human behaviours and crisis management. They employed mathematical models derived from kinetic theory and evolutionary game theory. In another study [24], crowd modelling was used to describe evacuation processes in constrained spaces, considering available data for decision-making. Multi-modal computer vision techniques, incorporating optical, thermal, and hyper-spectral imaging, were proposed for detecting crowd motions and behaviours in confined spaces such as Wang et al. [25] developed a model using latent path patterns to evaluate the fidelity of results compared to real-world situations, employing a Stochastic Variational Dual Hierarchical Dirichlet Process (SV-DHDP). Doulamis et al. [26] introduced a multi-disciplinary model for visual detection and tracking of crowd dynamics. Thermal imaging analysis was presented by [27], enabling localization and tracking of humans for evacuation purposes. Psychologists have proposed different models to depict human personality traits, such as the OCEAN and Psychoticism, Extraversion, and Neuroticism (PEN). Researchers have incorporated personality into crowd simulation to achieve more realistic behaviours including Durupinar et al. [28] mapped OCEAN traits to crowd behaviours, while Guy et al. [29] focused on generating heterogeneous crowd behaviours based on the PEN. Kim et al. [30] combined personality and adaptation syndrome theory for interactive crowd simulations. In emotion simulation, Ortony, Clore, and Collins (OCC), which focuses on appraisal, have been influential. Mehrabian's [31], Pleasure, Arousal, Dominance (PAD) distinguishes mood from emotion. Various studies have explored emotion contagion within groups, incorporated emotions into evacuation models, and utilized formal state-based techniques to model agents' levels.

## IV. EARLY EXPERIMENTS

Early NetLogo tool experimentation involve colouring particular patches to designate "safe" places inside the region. By mousing over them, agents could recognise these patches as "safe" places, and they were not permitted to cross them. The scenario represented severe weather events like typhoons, windstorms, or snowstorms. The evacuated agents are seen in the safe places in Fig. 1's depiction of the model state. Emotions and emotional contagion were not accounted for in the experimental model. Additionally, it lacked a transition from a panicked state to a non-panicked state.

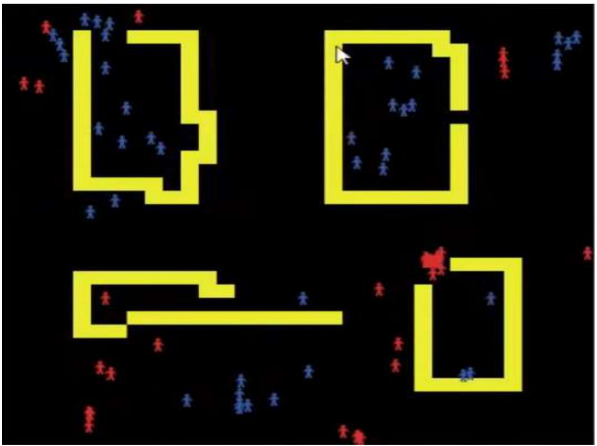

Fig. 1. Early Experiment

### A. Model Descriptions

Initialising the model using pre-defined places and obstacles is part of the process of creating the early model. The number of agents in the simulation is selectable by the user. Visual representations of how the emotional agent affects their evacuation process are the predicted outputs of the model. Panicked and non-panicked states are included in the model. Non-panicked states are represented by the colour green, alarmed states by the colour yellow, and panicked states by the colour red. The states are classified as reflex agents depending on the models. Every state has arbitrary values for emotional agents, mood, and personality. The personality does not alter throughout the evacuation event, although there shouldn't be any major shifts in the mood right away. The paper does not emphasise ongoing mood monitoring. Different occurrences, such as earthquakes, floods, and fires, have varying levels of danger, which has a bigger effect on the emotional agents nearby. The state's conduct is mostly influenced by the individual's personality and the emotional agent within the influence range. Mood determines the first emotional contagion, after which it regresses. It is suggested that the evacuation scenario be given a time constraint. Agents are deemed terminated if they don't leave the designated safe zones before the deadline.

## V. LATER EXPERIMENTS

The later model underwent major revisions after the early model description while keeping the fundamental concept. The switch from a pre-defined field to a user-designed field for the agents to work in was one significant revision. This choice sought to provide the model with more flexibility and set it apart from competing models. The emphasis was placed on mood and emotional condition rather than the idea of agent personality. This simplification was done to make room for the model's newly added ability to generate arbitrary fields. To more accurately reflect emotional states, the "non-panicked" state was divided into "alerted" and "calm" states. The model's initial state is indicated as "C-o" (Catastrophe-occurs), representing the initial intention to include environmental hazards or catastrophes in the model. As seen in Fig. 2, the agents transition between the states denoted by "pcitizen" (panicked citizen), "acitizen" (alerted citizen), and "ccitizen" (calm citizen). These citizen states can easily transition between each other because they are interchangeable. An agent can move on from these states by pressing "e-c" (evacuate) or "s-h" (seek help). The agent aggressively looks for an exit or help from others when in the "seek-help" state. The agent enters the "g-d" (get-directions) state if it is unable to flee on its own but comes into contact with a person of authority. In this state, it is given instructions on how to find the exit and help in the evacuation process. The dual-arrow transition between "seek-help" and "g-d" indicates that understanding the directions may require clarification, and the agent may request extra guidance. It is important to note that an agent can successfully flee and complete their mission without needing to meet a person of authority.

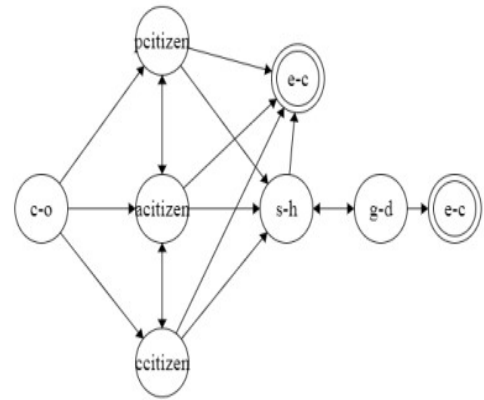

Fig. 2. Early Experiments States

### A. NetLogo

This section focuses on how the model was implemented using the agent-based modelling programme NetLogo. The section gives a succinct review and explanation of why NetLogo was chosen as the tool and environment. The NetLogo tool and the RePAST tool were the two primary tools initially taken into consideration. Other cutting-edge tools existed but were not used because of potential documentation and community support issues. Simulators, which are instruments for simulating model runs and computing models for analysis and education, come in a variety of shapes and sizes. Simulators for agent-based modelling include NetLogo and RePAST. But NetLogo sets itself apart by offering a unique programming language and a large library of resources, including a model library built from the Logo language. Updates since 2008 have concentrated on enhancing the NetLogo language, which is mostly written in Java. To boost performance, the compiler analyses, annotates, and restructures the user's code rather than directly interpreting it. Therefore, NetLogo was chosen.

## VI. NetLogo Simulations

The User Interface (UI) shown in Fig. 3 greets the user after the model has loaded. The UI offers a thorough overview of all its options and features. The UI has seven buttons, eight monitors, and one plot altogether. The agents, referred to as "turtles" in NetLogo, function in a setting made up of 3721 patches that measure 61*61. This subsection offers a thorough explanation of each UI options.

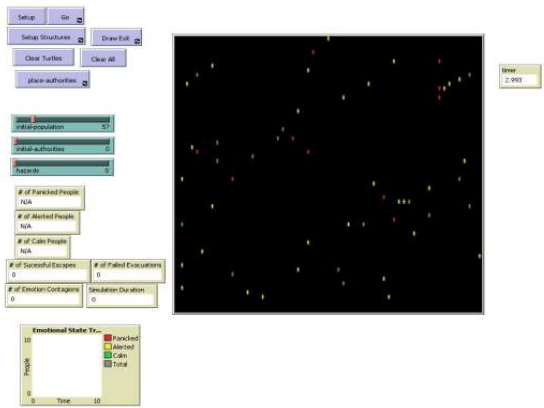

Fig. 3. GUI of NetLogo

### A. Breed/Variable Declarations

The breeds used in the program's code are shown in Fig. 4. A value that represents data of an unknown quantity is stored in a variable and is recognised by a symbolic name or identifier. Variables are crucial for storing and gaining access to the data required to carry out the logic of the programme. "Citizens" and "Authorities" are the two breeds represented in the model. The "Citizens" breed stands in for the turtles whose behaviour we hope to analyse, while the "Authorities" breed supports the turtles by helping them escape, much like in real-world situations. A breed of links was initially developed to keep track of instances of emotional contagion, but it was eventually abandoned because it could only build one link between two turtles, although in our scenario several contagions may take place. Now let's talk about the variables. NetLogo divides them into two categories: global variables and variables particular to turtles/patches/links. While variables declared for turtles (or other agent types) can only be accessed by those particular agents, whereas global variables can be accessed by all agents. Additionally, as shown in Fig. 5, it is feasible to declare variables for particular agent breeds.

The model's global variables are used to generate real-time data that is displayed to the user on the monitors. Except for "total citizens," which is used in the observer command centre to track the overall number of active citizens at any given time, all of these variables are particularly used for this purpose.

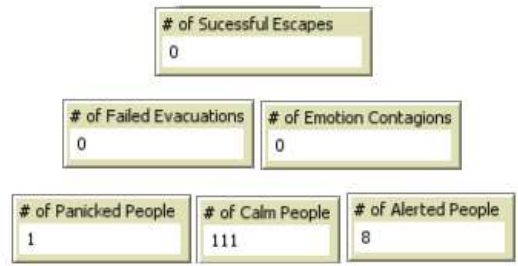

Fig. 4. Code Snippet of Breed/Variable Declarations

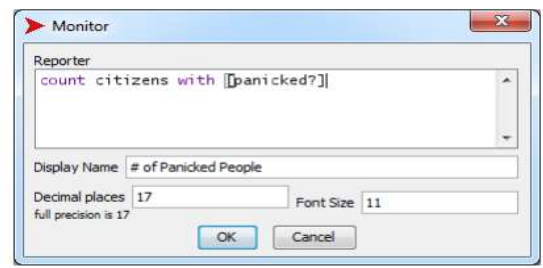

Fig. 5. Collection of GUI used to output data

The boolean variable "panicked?" "alerted?" and "calm?" are used by citizens to indicate whether they are in various emotional states. The corresponding numbers are displayed on monitors using these variables as well. Fig. 6 provides an instance of how this is accomplished. When looking at a turtle, the user can ignore the "turn-check" variable, an integer used for debugging. Its goal is to help the developer comprehend the route a turtle takes when coming across a barrier (grey patch). The "collision-check" process makes use of it. To help turtles evacuate, the "evacuation-directions?" variable, a boolean, is employed. Turtles behave slightly differently to find the exit when the value is true. The "get-directions" technique makes use of it. A turtle's mood is initially set using the "mood" variable, an integer that also tracks the turtle's mood as it moves through the programme.

Fig. 6. A Monitor's Code Window

## B. World Buildings

The user of the model can define their evacuation scenario using the offered code snippets in Fig. 7. By selecting "Setup Structures" and using the left mouse click, the "draw-building" option enables the user to paint grey patches that represent structures or wreckage. The drawing option is cancelled by pressing the button once more. Similarly, using the same option but using the colour magenta, the user can depict exits for the turtles to flee from. A simulation cannot begin without at least one exit (magenta patch), which is a key point to remember. A runtime error will appear if you try to perform it. The model can be stopped so the user can add more structures or exits. While pressing the "Clear All" option fully resets the world, pressing the "Clear Turtles" option enables the simulation to be repeated with the same world structure. The "place-authorities" option is just used for debugging and is not meant to be used in actual evacuation scenarios.

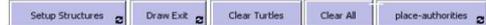

Fig. 7. Code Snippet of World Buildings

## C. Turtle Set-Up Initializations

The "setup-figures" option, which is a component of the offered snippet in Fig. 8, is in charge of initialising the variables and agent numbers for turtles and authorities around the world. At the start of each scenario, this option is used once. The sliders in the UI of the model can be used to change the desired numbers for turtles and authority. This option is made clearer, helps prevent bugs, and gives better control over the initialization process by being separated into a distinct procedure. The code at the end of the sample causes the "catastrophe-occurs" option to be executed. The "hazards" slider should not be utilised and should only be used for debugging, it is vital to remember.

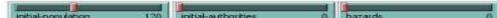

Fig. 8. Code Snippet of Turtle Set-Up Initializations

## D. Catastrophe Occurs

The turtles' emotional state is set to their default during the execution of the scenario by the "catastrophe-occurs" option. At the start of each scenario, it is run once. Although its name suggests a more general application, it is only employed for emotional initiation in this situation. It's vital to emphasise that even though authorities are also individuals, the paper does not focus on tracking their emotions. By looking at Fig. 9, it is possible to see that there is a 15% chance that a turtle will become frightened at the start of the scenario.

```
to catastrophe-occurs;  The "event" that initiates the whole process of evacuation
                        ;; It assigns statuses to all existing citizens and starts the evacuation

  set mood random 100

  ifelse mood >= 69
    [ calm-down ]
    [ if mood <= 68
      [ get-alerted ]
    ]

  if mood <= 15
    [get-panicked]

end
```

Fig. 9. Code Snippet of Catastrophe Occurs

## E. Agent Interactions

When turtles interact, the "interact" option is primarily in charge of simulating emotional contagion. It controls how turtles in various emotional states interact with one another. This option is shown in Fig. 10, where turtles with higher mood scores are more receptive to the effect of turtles with greater irritated mood scores. Based on their emotional condition and the emotional state of the other turtles they come into contact with, each turtle exhibits a particular behaviour. Consider a situation where three turtles come into contact with one another: one alert, one panicked and one calm. If they all come together at the same time, the "alerted" turtle's mood won't alter (-1 + 1 + 0). On the other hand, because the two other turtles (calm and alerted) do not affect the "panicked" turtle's demeanour, it will remain that way.

```
to interact; This procedure mimics emotional contagion
  ask citizens with [alerted?] [
    if any? citizens in-radius 2 with [panicked?]
      [ set mood mood - 1 ]
    if any? citizens in-radius 2 with [alerted?]
      [ set total-contagions total-contagions + 1 ]

    if any? citizens in-radius 2 with [calm?]
      [ set mood mood + 1 ]
    if any? other citizens in-radius 2 with [alerted?]
      [ set total-contagions total-contagions + 1 ]
  ]

  ask citizens with [calm?] [
    if any? citizens in-radius 2 with [alerted?]
      [ set mood mood - 1.5
        set total-contagions total-contagions + 1 ]
    if any? citizens in-radius 2 with [panicked?]
      [ set mood mood - 2
        set total-contagions total-contagions + 1 ]
    if any? other citizens in-radius 2 with [calm?]
      [ set mood mood + 0.5
        set total-contagions total-contagions + 1 ]
  ]

  ask citizens with [panicked?] [
    if any? other citizens in-radius 2 with [panicked?]
      [ set mood mood + 2
        set total-contagions total-contagions + 1 ]
    if any? citizens in-radius 2 with [alerted?]
      [ set total-contagions total-contagions + 1 ]
    if any? citizens in-radius 2 with [calm?]
      [ set total-contagions total-contagions + 1 ]
  ]
end
```

Fig. 10. Code Snippet of Agent Interactions

## VII. RESULT ANALYSIS

For the model's perspective uses, it is essential to evaluate its capabilities, especially in emergency situations. This section, which is in line with the main objective of the paper, concentrates only on modelling particular scenarios using representations from the real world. The model's performance is assessed while it is running, and the anticipated outcomes are shown in tables that are included with each scenario's description. Every scenario takes place in a 3721 patch, 61*61 patch universe. The exits take up a width of 10*5 patches (or 50 patches) and are always found at the edges of the world. For each scenario, a screenshot that complies with the requirements for that scenario is provided, showing how buildings and exits are distributed around the world. Additionally, different settings are used to conduct each scenario to simulate various evacuation scenarios. The population (or "citizen" turtles) and the existence of authorities (or "authority" turtles) are two examples of these criteria. Four variations of each scenario are made with different values for the population parameter to streamline the study. The variants come in three different numbers: "low population" with only 15 turtles, "medium population" with 75 turtles, and "high population" with 150 turtles. It is significant to note that a third criterion relating to risks like fires or floods was initially intended to be included but was eventually left out of the paper.

## A. Scenario 1

The first scenario examines sparsely populated regions, open spaces, or places devoid of structures for a variety of reasons. Additionally, it simulates enclosed settings like bars, theatres, and movie theatres. Buildings are purposefully left out of the simulation in this scenario in order to evaluate how well the agents function there. A graphic illustration of this situation is shown in Fig. 11.

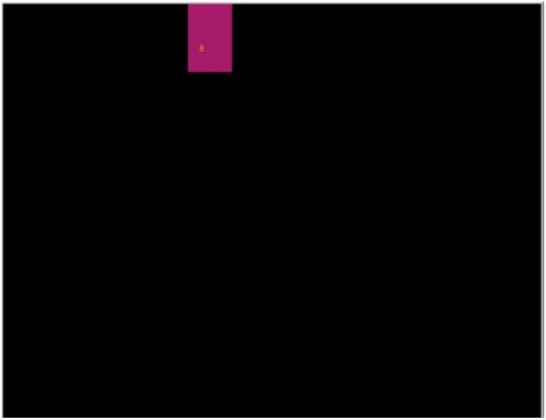

Fig. 11. Open Field Scenario

### 1) Low Population

A overview of the simulation attempts made in the paper is given in Table I (APPENDIX). It displays the number of successful and unsuccessful evacuations, emotional contagions, and the seconds in time of each evacuation effort. On average, 1.5 evacuations failed out of 13.7 simulation tries, whereas 68.5 emotional contagions happened. An evacuation lasted, on average, 120.6 seconds. During the simulated studies, most of the citizens (97.55%) were successfully evacuated.

### 2) Medium Population

The evaluation of a number of simulation attempts are summarised in Table II (APPENDIX). It offers information on the number of successful evacuations, unsuccessful evacuations, emotional contagions, and the seconds of time of each evacuation. On average, 5 evacuations failed to make it to the exit out of 65.6 simulation tries. During the evacuations, there were roughly 3623.5 emotional contagions, and each one lasted an average of 153.3 seconds. Approximately 90.93% of the evacuation attempts were successful, according to the results of the study. These results offer information about the model's performance and can be used to gauge how well it simulates emergency situations. Whereas Table 3 also presents the results of a series of simulation attempts with no authorities. In Table III (APPENDIX), on average, 9.6 evacuations didn't make it to the exit out of 63.4 simulated tries. During the evacuations, an average of 4012.5 emotional contagions were seen, and each one lasted an average of 184.7 seconds. The results show that 82.32% of the attempted evacuations were successful.

### 3) High Population

The numbers on successful evacuations, unsuccessful evacuations, emotional contagions, and the second of each evacuation in time units are presented in Table IV (APPENDIX) as the results of a series of simulation attempts. On average, 5 evacuations out of 114.3 simulation attempts failed to make it to the exit. During the evacuations, an average of 14,348.9 emotional contagions were seen, and each one lasted an average of 94.7 seconds. The results show that 68.77% of the attempted evacuations were successful. The results of ten simulation attempts are summarised in Table V (APPENDIX), which includes no authorities. Out of the ten simulation efforts, there were, on average, 109.1 attempts at evacuation, with about 40.9 successful evacuations and 68.2 unsuccessful ones. 15837.4 emotional contagions were recorded on average during the evacuations, and each evacuation lasted an average of 199.3 seconds. The time needed for the model to simulate the evacuation procedure is reflected in the average evacuation duration. In total, 62.51% of the citizens in these simulations were successfully evacuated.

## B. Scenario 2

The second scenario focuses on the evacuation of sparsely populated areas, including villages and rural settlements. It features three small buildings and a standard 10*5 exit, as depicted in Fig. 12.

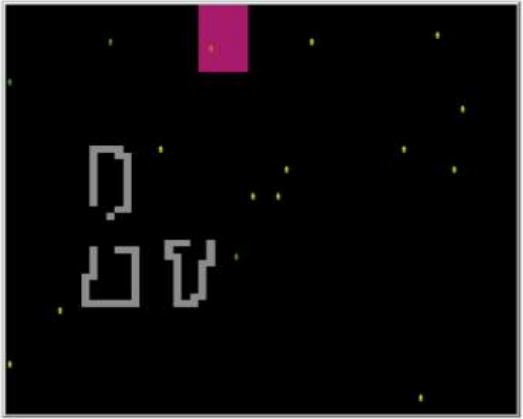

Fig. 12. Village Scenario

*1) Low Population*
The results of a simulation experiment including evacuations are shown in Table VI (APPENDIX). Ten different simulation attempts were made, each with a different scenario. The number of successful evacuations in various efforts ranged from 12 to 17. With the exception of two cases where two evacuations failed, none of the attempts at evacuation resulted in failures. In various tries, emotional contagions happened in the range of 21 to 262. The time required for the evacuation, expressed in seconds, ranged from 35 to 619. The simulation had a success rate of 97.55% on average, meaning that most evacuations were successful. The differences in emotional contagions and evacuation times between attempts show the impact of many elements on the evacuation process.

*2) Medium Population*
The results of a simulation experiment on evacuations are shown in Table VII (APPENDIX). The simulation was run ten times, and the average results showed that 61 to 71 successful evacuations occurred on average, with a success rate of 68.2. Some evacuation attempts failed, with an average of 8.2 and a range of 1 to 13. The occurrence of emotional contagions ranged from 3041 to 4144, with an average of 3472.1. With an average of 149.2, the evacuation time, expressed in seconds, ranged from 77 to 220. The simulation had an average success rate of 88.05%, which denotes a respectably good success rate for evacuations. The complex nature of the evacuation process is shown by the existence of unsuccessful evacuations and a range of emotional contagions. The results of a simulation experiment on evacuations with no authorities are summarised in Table VIII (APPENDIX). The number of successful evacuations, which ranged from 60 to 72, was 67.1 on average. The number of unsuccessful evacuation efforts varied from 3 to 15, with an average of 7.1. The number of emotional contagions ranged from 3438 to 7773, with an average of 4842. In seconds, the evacuation time ranged from 87 to 404, with an average of 182. A reasonably high success percentage in evacuations is indicated by the simulation's average success rate of 87.89%.

*3) High Population*
The results of a simulation experiment on evacuations are shown in Table IX (APPENDIX). With an average of 116.6, the number of successful evacuations ranged from 103 to 130.

There were several unsuccessful evacuation attempts, with an average of 33.4 and a range of 20 to 47. 12,917 to 18,117 cases of emotional contagion were reported, with a mean of 15,615.7. Seconds representing the length of the evacuation ranged from 23 to 257, with 117.6 being the average. The simulation had a success percentage of 71.46% on average, which denotes a decent success in evacuations. The difficulties and complexity of the evacuation procedure are highlighted by the occurrence of unsuccessful evacuations and various emotional contagions. The seconds of evacuations also varied, indicating the impact of several factors on the total amount of time needed. Data from a simulation experiment on evacuations with no authorities are presented in Table X (APPENDIX). There were 111.6 simulation attempts, ranging in size from 95 to 124. There were 26 to 55 successful evacuations, averaging 38.4. Failures ranged from 14,304 to 19,691, with an average of 16,649. The evacuation time, expressed in seconds, ranged from 84 to 145, with 134.1 being the average. Overall, the simulation showed a moderate average success rate with various levels of evacuation success.

*C. Scenario 3*
The purpose of the third scenario is to imitate medium-sized towns and collect data in accordance. As seen in Fig. 13, it incorporates six buildings and the common 10*5 exit.

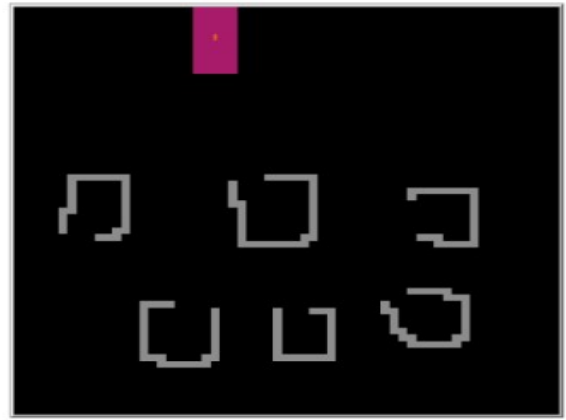

Fig. 13. Town Scenario

*1) Low Population*
The average number of successful evacuations in a series of simulation trials, as shown in Table XI (APPENDIX), was 14.4, whereas the average number of unsuccessful evacuations was 0.7. The average number of emotional contagions was 206.3, while the typical length of an evacuation was 156.9 seconds.

*2) Medium Population*
In a number of simulation attempts, as depicted in Table XII (APPENDIX), the average number of successful evacuations was 66.6, whereas the average number of unsuccessful evacuations was 8.6. The average number of emotional contagions was 4,376.8, and the typical length of an evacuation was 128 seconds. As indicated in Table XIII (APPENDIX), in a number of simulated attempts without authorities, the average number of successful evacuations was 68.7, whereas the average number of unsuccessful evacuations was 6.3. The average number of emotional contagions was 5,245.4, and the typical length of each evacuation was 190.6 seconds.

### 3) High Population

The average number of successful evacuations in a series of simulation attempts, as shown in Table XIV (APPENDIX), was 115.2, while the average number of unsuccessful evacuations was 34.8. The average number of emotional contagions was 16,419.7, and the typical length of an evacuation was 220.5 seconds. As indicated in Table XV (APPENDIX), in a series of simulated attempts without authorities, the average number of successful evacuations was 106.4, whereas the average number of unsuccessful evacuations was 43.6. The average number of emotional contagions was 18,904.4, and the typical length of an evacuation was 164.6 seconds.

### D. Scenario 4

The "city" scenario, which simulates major cities, capitals, and other metropolitan areas, is the last scenario. As shown in Fig. 14, it has nine buildings and a patched-together conventional 10*5 exit.

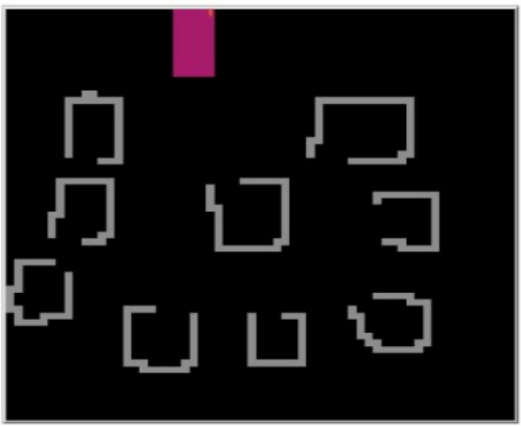

Fig. 14. City Scenario

### 1) Low Population

In a series of simulation attempts, as depicted in Table XVI (APPENDIX), the average number of successful evacuations was 14.4, whereas the average number of unsuccessful evacuations was 1.1. The average number of emotional contagions was 298, and the typical length of an evacuation was 265.6 seconds.

### 2) Medium Population

The average number of successful evacuations in a series of simulation trials, as indicated in Table XVII (APPENDIX), was 66, whereas the average number of unsuccessful evacuations was 9. The average number of emotional contagions was 5,084.4, and the typical evacuation time was 199.6 seconds. The average number of successful evacuations in a series of simulation attempts without the involvement of authorities as shown in Table XVII (APPENDIX) is 65.2, whereas the average number of unsuccessful evacuations is 9.8. The average number of emotional contagions was 7,048.5, and the typical length of each evacuation was 217.7 seconds.

### 3) High Population

The average number of successful evacuations in a series of simulation attempts, as indicated in Table XIX (APPENDIX), was 108.5, while the average number of unsuccessful evacuations was 41.5. The average number of emotional contagions was 17,184.2, and the typical evacuation time was 231.9 seconds. As indicated in Table XX (APPENDIX), in a series of simulated attempts with no authorities, the average number of successful evacuations was 107.8, whereas the average number of unsuccessful evacuations was 44.2. The average number of emotional contagions was 21,296.3, and the typical evacuation time was 199.8 seconds.

### E. Overall Results

Tables XXI to XXIV (APPENDIX) provide easy access to the results, these tables serve to summarise the results and make it easier to evaluate the model's overall capabilities. Data for several evacuation scenarios based on varied population sizes and the presence or absence of authorities are presented in Tables XXI to XXIV (APPENDIX). Table XXI demonstrates that the low population have been 14.2 successful evacuations on average, while only 0.2 have failed. The duration of the evacuation is 120.8 seconds, and the emotional contagions are mild (68.2). Medium population with four authorities when compared to the low population, this scenario exhibits a significantly higher number of successful evacuations (69.1). With an evacuation length of 151.5 seconds, failed evacuations and emotional contagions both marginally rise. No authorities and a medium-sized population show similar results to the previous scenario, however, there are still a lot of successful evacuations (62.5), but there are also more failed evacuations and emotional contagions. There are 184.7 seconds in the evacuation time. With four authorities and a large population shows noticeable increase in unsuccessful evacuations and emotional contagions, whereas the number of successful evacuations continues to rise to 114.1. The time of the evacuation, nevertheless, is cut in half to 94.7 seconds. Numerous people but no authorities show that the average number of successful evacuations in this scenario is the highest (109.9), but there is also a noticeable rise in unsuccessful evacuations and emotional contagions. There are 199.3 seconds in the evacuation time. While Table XXII displays the results of evacuation attempts and emotional contagion in various situations based on the population of the city and the existence or absence of authorities. Four authorities and a low to medium population shows an evacuation duration of 196.2 seconds, successful evacuations of 14.1k, unsuccessful evacuations of 0.1k, and an emotional contagion of 116.2k. Medium population/no authorities shows 67.4k successful evacuations, 7.3k unsuccessful evacuations, 3473.8k emotional contagions, and 148.1 seconds for the duration of the evacuation. Medium population/four authorities shows an emotional contagion of 4482k; successful evacuations of 66.2k; failed evacuations of 8.2k; evacuation duration: 182 seconds. High population/no authorities shows 116.7k successful evacuations, 33.7k unsuccessful evacuations, 15617.7k emotional contagions, and 117.1 seconds of actual evacuation time. High population/four authorities shows evacuations successfully completed with 111.6k; unsuccessfully completed: 38.4k; emotional contagion: 16649.4k; time of evacuation: 134.1 seconds. In a town with diverse demographics and authority, effective evacuations ranged from 14.9 to 115.2, whereas failed evacuations ranged from 0.5 to 43.6, as shown in Tables XXIII and XXIV. There were varying degrees of emotional contagions, ranging in intensity from 207.2 to 18,904.4. The length of the evacuation ranged from 128 to 269.1 seconds. Whereas in Table XXIV ranged from 14.6 to 108.5, whereas unsuccessful ones ranged from 0.6 to 43.1. There were varying degrees of emotional contagions, ranging in intensity from 300.6 to 21,296.3. The length of the evacuation ranged from 199.6 to 262.6 seconds.

According to the research results, although the improvement is not great, the presence of authorities does increase the possibility that individuals will leave a building. Additionally, in each iteration of the scenario, the incidence of emotional contagions among the turtles rises as the population size does. It is also clear that higher population densities result in higher casualty rates, which can be explained by the turtles' relatively random movement as they look for an exit. Results from this experiment are sufficient for the intended scope, while more exact results might be obtained with a bigger sample.

## VIII. CONCLUSIONS AND FUTURE WORKS

While the paper was successful in achieving its objectives, it should be noted that the model is not currently suitable for real-life applications. Nonetheless, there are opportunities to further enhance the model. One area of improvement is expanding the features of the model. For instance, additional monitors could be implemented to capture the states of evacuated turtles, such as whether they are alerted or in a panicked state. These monitors would provide valuable data for statistical analysis and further understanding of the evacuation process. The existing procedures within the model can be leveraged to facilitate the integration of these monitors. Furthermore, the model could be extended to include environmental hazards, as initially discussed in the system description. These hazards could act as deterrents for the turtles in their vicinity, influencing their emotional states and posing risks to their well-being during the evacuation process. By incorporating these hazards, the model would better simulate real-life emergency situations and provide insights into the dynamics of evacuations in the presence of environmental challenges. So, there is ample scope for future work and enhancements to the model. By incorporating additional features such as monitors for tracking evacuated turtles and introducing environmental hazards, the model can be further refined and offer a more comprehensive simulation of emergency evacuations. These improvements would contribute to a deeper understanding of evacuation dynamics and assist in the development of more effective strategies for real-life emergency situations.

## IX. DECLARATIONS

A. **Funding:** No funds, grants, or other support was received.

B. **Conflict of Interest:** The authors declare that they have no known competing for financial interests or personal relationships that could have appeared to influence the work reported in this paper.

C. **Data Availability:** Data will be made on reasonable request.

D. **Code Availability:** Code will be made on reasonable request.

APPENDIX

**TABLE I**
**LOW POPULATION (15) WITH NO AUTHORITIES (0)**

| Simulation Attempts | Successful Evacuations | Failed Evacuations | Emotional Contagions | Evacuation Duration |
|---|---|---|---|---|
| #1 | 13 | 2 | 83 | 192 |
| #2 | 11 | 4 | 75 | 63 |
| #3 | 16 | 0 | 32 | 55 |
| #4 | 14 | 0 | 95 | 191 |
| #5 | 16 | 1 | 72 | 182 |
| #6 | 14 | 0 | 93 | 78 |
| #7 | 13 | 2 | 81 | 143 |
| #8 | 17 | 0 | 57 | 52 |
| #9 | 12 | 4 | 43 | 92 |
| #10 | 11 | 2 | 54 | 158 |
| Average (mean) | 13.7 | 1.5 | 68.5 | 120.6 |
| | 97.55% of Citizens | | | |

TABLE II
MEDIUM POPULATION (75) WITH FOUR AUTHORITIES (4)

| Simulation Attempts | Successful Evacuations | Failed Evacuations | Emotional Contagions | Evacuation Duration |
|---|---|---|---|---|
| #1 | 72 | 4 | 4401 | 206 |
| #2 | 62 | 8 | 3701 | 346 |
| #3 | 72 | 1 | 3451 | 76 |
| #4 | 62 | 7 | 3431 | 96 |
| #5 | 72 | 1 | 4051 | 196 |
| #6 | 62 | 9 | 3142 | 47 |
| #7 | 62 | 7 | 3163 | 127 |
| #8 | 62 | 5 | 3914 | 227 |
| #9 | 62 | 4 | 3125 | 169 |
| #10 | 68 | 4 | 3856 | 43 |
| Average (mean) | 65.6 | 5 | 3623.5 | 153.3 |
| | 90.93% | | | |

TABLE III
MEDIUM POPULATION (75) WITH NO AUTHORITIES (0)

| Simulation Attempts | Successful Evacuations | Failed Evacuations | Emotional Contagions | Evacuation Duration |
|---|---|---|---|---|
| #1 | 64 | 16 | 2833 | 98 |
| #2 | 64 | 6 | 4665 | 82 |
| #3 | 64 | 6 | 4839 | 35 |
| #4 | 64 | 8 | 3496 | 478 |
| #5 | 71 | 4 | 4815 | 183 |
| #6 | 62 | 12 | 4052 | 346 |
| #7 | 62 | 11 | 3646 | 134 |
| #8 | 62 | 7 | 3892 | 163 |
| #9 | 63 | 9 | 4120 | 150 |
| #10 | 53 | 12 | 3767 | 178 |
| Average (mean) | 63.4 | 9.6 | 4012.5 | 184.7 |
| | 82.32% | | | |

**TABLE IV**
**HIGH POPULATION (150) WITH FOUR AUTHORITIES (4)**

| Simulation Attempts | Successful Evacuations | Failed Evacuations | Emotional Contagions | Evacuation Duration |
|---|---|---|---|---|
| #1 | 111 | 39 | 13208 | 98 |
| #2 | 104 | 46 | 15920 | 121 |
| #3 | 102 | 48 | 12169 | 88 |
| #4 | 122 | 28 | 15679 | 103 |
| #5 | 124 | 26 | 15069 | 84 |
| #6 | 128 | 22 | 14998 | 145 |
| #7 | 114 | 36 | 14484 | 63 |
| #8 | 115 | 35 | 14173 | 85 |
| #9 | 109 | 41 | 14022 | 58 |
| #10 | 114 | 36 | 13767 | 102 |
| Average (mean) | 114.3 | 35.7 | 14348.9 | 94.7 |
| | 68.77% | | | |

**TABLE V**
**HIGH POPULATION (150) WITH NO AUTHORITIES (0)**

| Simulation Attempts | Successful Evacuations | Failed Evacuations | Emotional Contagions | Evacuation Duration |
|---|---|---|---|---|
| #1 | 107 | 43 | 17487 | 174 |
| #2 | 114 | 36 | 14780 | 117 |
| #3 | 103 | 47 | 13626 | 241 |
| #4 | 113 | 37 | 14657 | 205 |
| #5 | 118 | 32 | 16247 | 106 |
| #6 | 113 | 37 | 20478 | 217 |
| #7 | 111 | 39 | 15838 | 146 |
| #8 | 104 | 46 | 14766 | 76 |
| #9 | 107 | 43 | 15255 | 343 |
| #10 | 101 | 49 | 15240 | 368 |
| Average (mean) | 109.1 | 40.9 | 15837.4 | 199.3 |
| | 62.51% | | | |

TABLE VI
LOW POPULATION (15) WITH NO AUTHORITIES (0)

| Simulation Attempts | Successful Evacuations | Failed Evacuations | Emotional Contagions | Evacuation Duration |
|---|---|---|---|---|
| #1 | 12 | 0 | 262 | 82 |
| #2 | 12 | 0 | 153 | 619 |
| #3 | 12 | 2 | 182 | 112 |
| #4 | 13 | 0 | 21 | 104 |
| #5 | 13 | 0 | 37 | 207 |
| #6 | 16 | 2 | 50 | 168 |
| #7 | 16 | 2 | 139 | 35 |
| #8 | 16 | 2 | 40 | 150 |
| #9 | 17 | 2 | 22 | 188 |
| #10 | 17 | 0 | 260 | 311 |
| Average (mean) | 16.4 | 0.5 | 117.9 | 197.2 |
| | 97.55% | | | |

TABLE VII

**MEDIUM POPULATION (75) WITH FOUR AUTHORITIES (4)**

| Simulation Attempts | Successful Evacuations | Failed Evacuations | Emotional Contagions | Evacuation Duration |
|---|---|---|---|---|
| #1 | 62 | 13 | 3421 | 77 |
| #2 | 62 | 9 | 3041 | 220 |
| #3 | 62 | 11 | 3092 | 208 |
| #4 | 71 | 3 | 3962 | 126 |
| #5 | 61 | 11 | 3273 | 82 |
| #6 | 71 | 4 | 3543 | 134 |
| #7 | 71 | 1 | 3304 | 151 |
| #8 | 61 | 9 | 4144 | 215 |
| #9 | 65 | 13 | 3415 | 172 |
| #10 | 65 | 8 | 3525 | 99 |
| Average (mean) | 68.2 | 8.2 | 3472.1 | 149.2 |
| | 88.05% | | | |

**TABLE VIII**
**MEDIUM POPULATION (75) WITH NO AUTHORITIES (0)**

| Simulation Attempts | Successful Evacuations | Failed Evacuations | Emotional Contagions | Evacuation Duration |
|---|---|---|---|---|
| #1 | 67 | 9 | 4631 | 220 |
| #2 | 67 | 12 | 3891 | 118 |
| #3 | 62 | 8 | 7773 | 230 |
| #4 | 72 | 3 | 5393 | 250 |
| #5 | 66 | 9 | 4405 | 138 |
| #6 | 67 | 8 | 4637 | 95 |
| #7 | 60 | 15 | 3438 | 87 |
| #8 | 68 | 7 | 6301 | 154 |
| #9 | 70 | 5 | 3606 | 114 |
| #10 | 70 | 5 | 4254 | 404 |
| Average (mean) | 67.1 | 7.1 | 4842 | 182 |
| | 87.89% | | | |

**TABLE IX**
**HIGH POPULATION (150) WITH FOUR AUTHORITIES (4)**

| Simulation Attempts | Successful Evacuations | Failed Evacuations | Emotional Contagions | Evacuation Duration |
|---|---|---|---|---|
| #1 | 130 | 20 | 18117 | 172 |
| #2 | 114 | 36 | 12931 | 28 |
| #3 | 106 | 44 | 18047 | 257 |
| #4 | 109 | 41 | 17156 | 75 |
| #5 | 125 | 25 | 15679 | 23 |
| #6 | 114 | 36 | 13421 | 70 |
| #7 | 125 | 25 | 17096 | 124 |
| #8 | 103 | 47 | 14168 | 133 |
| #9 | 116 | 34 | 14522 | 131 |
| #10 | 124 | 26 | 15020 | 163 |
| Average (mean) | 116.6 | 33.4 | 15615.7 | 117.6 |
| | 71.46% | | | |

**TABLE X**
**HIGH POPULATION (150) WITH NO AUTHORITIES (0)**

| Simulation Attempts | Successful Evacuations | Failed Evacuations | Emotional Contagions | Evacuation Duration |
|---|---|---|---|---|
| #1 | 111 | 39 | 16378 | 278 |
| #2 | 124 | 26 | 17965 | 97 |
| #3 | 95 | 55 | 14629 | 96 |
| #4 | 114 | 36 | 19691 | 118 |
| #5 | 112 | 38 | 16453 | 142 |
| #6 | 112 | 38 | 16158 | 84 |
| #7 | 115 | 35 | 17103 | 122 |
| #8 | 109 | 41 | 14304 | 145 |
| #9 | 113 | 37 | 17021 | 128 |
| #10 | 111 | 39 | 16792 | 131 |
| Average mean) | 111.6 | 38.4 | 16649.4 | 134.1 |

**TABLE XI**
**LOW POPULATION (15) WITH NO AUTHORITIES (0)**

| Simulation Attempts | Successful Evacuations | Failed Evacuations | Emotional Contagions | Evacuation Duration |
|---|---|---|---|---|
| #1 | 12 | 0 | 31 | 143 |
| #2 | 12 | 2 | 21 | 154 |
| #3 | 13 | 0 | 71 | 432 |
| #4 | 13 | 0 | 571 | 131 |
| #5 | 14 | 0 | 451 | 84 |
| #6 | 14 | 2 | 101 | 223 |
| #7 | 16 | 0 | 41 | 93 |
| #8 | 16 | 2 | 482 | 143 |
| #9 | 17 | 0 | 112 | 93 |
| #10 | 17 | 1 | 182 | 73 |
| Average (mean) | 14.4 | 0.7 | 206.3 | 156.9 |
| | 97.96% | | | |

**TABLE XII**
**MEDIUM POPULATION (75) WITH FOUR AUTHORITIES (4)**

| Simulation Attempts | Successful Evacuations | Failed Evacuations | Emotional Contagions | Evacuation Duration |
|---|---|---|---|---|
| #1 | 72 | 3 | 4203 | 170 |
| #2 | 62 | 12 | 2943 | 151 |
| #3 | 62 | 6 | 5624 | 212 |
| #4 | 72 | 6 | 4246 | 77 |
| #5 | 73 | 7 | 5389 | 82 |
| #6 | 63 | 12 | 6324 | 79 |
| #7 | 63 | 8 | 4451 | 80 |
| #8 | 73 | 9 | 5046 | 120 |
| #9 | 63 | 12 | 3205 | 92 |
| #10 | 63 | 11 | 2337 | 217 |
| Average (mean) | 66.6 | 8.6 | 4376.8 | 128 |
| | 88.72% | | | |

**TABLE XIII**
**MEDIUM POPULATION (75) WITH NO AUTHORITIES (0)**

| Simulation Attempts | Successful Evacuations | Failed Evacuations | Emotional Contagions | Evacuation Duration |
|---|---|---|---|---|
| #1 | 67 | 8 | 5725 | 108 |
| #2 | 67 | 8 | 6007 | 434 |
| #3 | 72 | 3 | 4572 | 165 |
| #4 | 73 | 2 | 4550 | 105 |
| #5 | 66 | 9 | 6935 | 235 |
| #6 | 67 | 8 | 4762 | 258 |
| #7 | 65 | 10 | 3830 | 276 |
| #8 | 67 | 8 | 4460 | 80 |
| #9 | 72 | 3 | 5892 | 93 |
| #10 | 71 | 4 | 5721 | 152 |
| Average (mean) | 68.7 | 6.3 | 5245.4 | 190.6 |
| | 92.87% | | | |

**TABLE XIV**
**HIGH POPULATION (150) WITH FOUR AUTHORITIES (4)**

| Simulation Attempts | Successful Evacuations | Failed Evacuations | Emotional Contagions | Evacuation Duration |
|---|---|---|---|---|
| #1 | 110 | 40 | 14322 | 92 |
| #2 | 115 | 35 | 14566 | 148 |
| #3 | 118 | 32 | 15142 | 95 |
| #4 | 112 | 38 | 19476 | 38 |
| #5 | 113 | 37 | 15625 | 222 |
| #6 | 110 | 40 | 16854 | 69 |
| #7 | 119 | 31 | 15900 | 920 |
| #8 | 117 | 33 | 17917 | 171 |
| #9 | 116 | 34 | 15254 | 98 |
| #10 | 122 | 28 | 19141 | 352 |
| Average (mean) | 115.2 | 34.8 | 16419.7 | 220.5 |
| | 69.79% | | | |

**TABLE XV**
**HIGH POPULATION (150) WITH NO AUTHORITIES (0)**

| Simulation Attempts | Successful Evacuations | Failed Evacuations | Emotional Contagions | Evacuation Duration |
|---|---|---|---|---|
| #1 | 104 | 46 | 26380 | 303 |
| #2 | 112 | 38 | 20991 | 210 |
| #3 | 103 | 47 | 15670 | 90 |
| #4 | 100 | 50 | 15479 | 190 |
| #5 | 109 | 41 | 18699 | 134 |
| #6 | 102 | 48 | 15987 | 81 |
| #7 | 116 | 34 | 26642 | 390 |
| #8 | 112 | 38 | 19192 | 98 |
| #9 | 103 | 47 | 13500 | 80 |
| #10 | 103 | 47 | 16504 | 70 |
| Average (mean) | 106.4 | 43.6 | 18904.4 | 164.6 |
| | 59.02% | | | |

**TABLE XVI**
**LOW POPULATION (15) WITH NO AUTHORITIES (0)**

| Simulation Attempts | Successful Evacuations | Failed Evacuations | Emotional Contagions | Evacuation Duration |
|---|---|---|---|---|
| #1 | 14 | 0 | 561 | 747 |
| #2 | 14 | 0 | 401 | 207 |
| #3 | 13 | 2 | 481 | 77 |
| #4 | 13 | 0 | 62 | 137 |
| #5 | 16 | 0 | 413 | 347 |
| #6 | 16 | 0 | 342 | 147 |
| #7 | 17 | 3 | 412 | 96 |
| #8 | 17 | 0 | 102 | 286 |
| #9 | 12 | 2 | 142 | 136 |
| #10 | 12 | 4 | 71 | 476 |
| Average (mean) | 14.4 | 1.1 | 298.7 | 265.6 |
| | 96.55% | | | |

**TABLE XVII**
**MEDIUM POPULATION (75) WITH FOUR AUTHORITIES (4)**

| Simulation Attempts | Successful Evacuations | Failed Evacuations | Emotional Contagions | Evacuation Duration |
|---|---|---|---|---|
| #1 | 65 | 10 | 6876 | 138 |
| #2 | 63 | 12 | 5918 | 112 |
| #3 | 66 | 9 | 3469 | 201 |
| #4 | 68 | 7 | 5529 | 385 |
| #5 | 69 | 6 | 4148 | 231 |
| #6 | 67 | 8 | 6077 | 310 |
| #7 | 63 | 12 | 3246 | 126 |
| #8 | 68 | 7 | 5975 | 154 |
| #9 | 68 | 7 | 4938 | 168 |
| #10 | 63 | 12 | 4668 | 171 |
| Average (mean) | 66 | 9 | 5084.4 | 199.6 |
| | 86.36% | | | |

**TABLE XVIII**
**MEDIUM POPULATION (75) WITH NO AUTHORITIES (0)**

| Simulation Attempts | Successful Evacuations | Failed Evacuations | Emotional Contagions | Evacuation Duration |
|---|---|---|---|---|
| #1 | 70 | 5 | 5604 | 181 |
| #2 | 66 | 9 | 6067 | 175 |
| #3 | 66 | 9 | 4947 | 214 |
| #4 | 62 | 13 | 12367 | 293 |
| #5 | 67 | 8 | 9715 | 275 |
| #6 | 59 | 16 | 8796 | 192 |
| #7 | 64 | 11 | 6168 | 158 |
| #8 | 69 | 6 | 7268 | 483 |
| #9 | 66 | 9 | 5802 | 84 |
| #10 | 63 | 12 | 3751 | 122 |
| Average mean) | 65.2 | 9.8 | 7048.5 | 217.7 |
| | 84.97% | | | |

**TABLE XIX**
**HIGH POPULATION (150) WITH FOUR AUTHORITIES (4)**

| Simulation Attempts | Successful Evacuations | Failed Evacuations | Emotional Contagions | Evacuation Duration |
|---|---|---|---|---|
| #1 | 100 | 50 | 14071 | 195 |
| #2 | 110 | 40 | 17295 | 124 |
| #3 | 108 | 42 | 18369 | 78 |
| #4 | 109 | 41 | 17707 | 149 |
| #5 | 116 | 34 | 16158 | 609 |
| #6 | 102 | 48 | 17839 | 198 |
| #7 | 110 | 40 | 17481 | 140 |
| #8 | 112 | 38 | 16625 | 282 |
| #9 | 116 | 34 | 17188 | 315 |
| #10 | 102 | 48 | 19109 | 229 |
| Average (mean) | 108.5 | 41.5 | 17184.2 | 231.9 |
| | 61.75% | | | |

**TABLE XX**
**HIGH POPULATION (150) WITH NO AUTHORITIES (0)**

| Simulation Attempts | Successful Evacuations | Failed Evacuations | Emotional Contagions | Evacuation Duration |
|---|---|---|---|---|
| #1 | 101 | 55 | 25143 | 158 |
| #2 | 111 | 35 | 20005 | 203 |
| #3 | 101 | 45 | 25806 | 184 |
| #4 | 111 | 35 | 20683 | 312 |
| #5 | 101 | 45 | 18806 | 122 |
| #6 | 102 | 46 | 18091 | 140 |
| #7 | 112 | 46 | 19127 | 106 |
| #8 | 112 | 37 | 22926 | 225 |
| #9 | 92 | 57 | 21983 | 157 |
| #10 | 102 | 57 | 20393 | 391 |
| Average (mean) | 107.8 | 44.2 | 21296.3 | 199.8 |
| | 59.68% | | | |

**TABLE XXI**
**SCENARIO 1**

| Scenario #1."OpenField" | Low Population | Medium Population/Four authorities | Medium Population/ No Authorities | High Population/Four Authorities | High Population/ No Authorities |
|---|---|---|---|---|---|
| Successful Evacuations | 14.2 | 69.1 | 62.5 | 114.1 | 109.9 |
| Failed Evacuations | 0.2 | 5.1 | 9.3 | 35.2 | 40.6 |
| Emotional Contagions | 68.2 | 3621.4 | 4011.2 | 14348.1 | 15837.3 |
| Evacuation Duration | 120.8 | 151.5 | 184.7 | 94.7 | 199.3 |

**TABLE XXII**

**SCENARIO 2**

| Scenario #2."City" | Low Population | Medium Population/Four Authorities | Medium Population/ No Authorities | High Population/Four Authorities | High Population/ No Authorities |
|---|---|---|---|---|---|
| Successful Evacuations | 14.1 | 67.4 | 66.2 | 116.7 | 111.6 |
| Failed Evacuations | 0.1 | 7.3 | 8.2 | 33.7 | 38.4 |
| Emotional Contagions | 116.2 | 3473.8 | 4482 | 15617.7 | 16649.4 |
| Evacuation Duration | 196.2 | 148.1 | 182 | 117.1 | 134.1 |

**TABLE XXIII**
**SCENARIO 3**

| Scenario #3. "Town" | Low Population | Medium Population/Four Authorities | Medium Population/ No Authorities | High Population/Four Authorities | High Population/ No Authorities |
|---|---|---|---|---|---|
| Successful Evacuations | 14.9 | 68.7 | 67.4 | 115.2 | 106.4 |
| Failed Evacuations | 0.5 | 7.6 | 6.3 | 34.8 | 43.6 |
| Emotional Contagions | 207.2 | 4376.8 | 5245.4 | 16419.7 | 18904.4 |
| Evacuation Duration | 269.1 | 128 | 190.6 | 220.5 | 164.6 |

**TABLE XXIV**
**SCENARIO 4**

| Scenario #4. "City" | Low Population | Medium Population/Four Authorities | Medium Population/ No Authorities | High Population/Four Authorities | High Population/ No Authorities |
|---|---|---|---|---|---|
| Successful Evacuations | 14.6 | 66 | 65.2 | 108.5 | 106.9 |
| Failed Evacuations | 0.6 | 9 | 9.8 | 41.5 | 43.1 |
| Emotional Contagions | 300.6 | 5084.4 | 7048.5 | 17184.2 | 21296.3 |
| Evacuation Duration | 262.6 | 199.6 | 217.7 | 231.9 | 199.8 |